\documentclass[%
aip,
prfluids,%
amsmath,amssymb,
preprint,%
]{revtex4-1}
\usepackage{graphicx}
\usepackage[utf8]{inputenc}
\usepackage{amsmath}
\usepackage{natbib}
\usepackage{xcolor}
\usepackage{hyperref}
\usepackage{cleveref}
\usepackage{ulem}

\graphicspath{{final_figs/}}

\newcommand{\pdts}[1]{\partial_t #1}
\newcommand{\pdt}[1]{\partial_t\left(#1\right)}

\newcommand{\pdx}[1]{\partial_x\left(#1\right)}
\newcommand{\pdks}[1]{\partial_k #1}
\newcommand{\pdk}[1]{\partial_k\left(#1\right)}
\newcommand{\pdis}[1]{\partial_i #1}

\newcommand{\pdjs}[1]{\partial_j #1}
\newcommand{\pdj}[1]{\partial_j\left(#1\right)}
\newcommand{\R}{\tilde{R}}

\newcommand{\alp}{\alpha_\mathrm{p}}
\newcommand{\alps}{\alpha_\mathrm{sep}}
\newcommand{\Dp}{D_\mathrm{p}}
\newcommand{\Rep}{\mathrm{Re}_\mathrm{p}}
\newcommand{\CDx}{C_{\mathrm{D},x}}
\newcommand{\M}{\mathrm{Ma}}

\newcommand{\tauL}{\tau_L}
\newcommand{\phrho}{\phavg{\rho}}

\newcommand{\php}{\phavg{p}}

\newcommand{\phavg}[1]{\langle #1 \rangle}
\newcommand{\favg}[1]{\tilde{#1}}
\newcommand{\wfavg}[1]{\widetilde{#1}}

\newcommand{\fu}{\favg{u}}

\begin{document}

\title{Particle-resolved simulations of shock-induced flow through particle clouds at different Reynolds numbers}

\author{Andreas Nygård Osnes}
\email{a.n.osnes@its.uio.no}
\affiliation{Department of Technology Systems, University of Oslo, P.O. Box 70 Kjeller, NO-2027 Kjeller, Norway}

\author{Magnus Vartdal}
\email{Magnus.Vartdal@ffi.no}
\affiliation{Norwegian Defence Research Establishment, P.O. Box 25, NO-2027 Kjeller, Norway}

\author{Marianne Gjestvold Omang}
\email{m.g.omang@astro.uio.no}
\affiliation{Norwegian Defence Estates Agency, P.O. Box 405 Sentrum, NO-0103 Oslo, Norway}
\affiliation{Institute of Theoretical Astrophysics, University of Oslo, P.O. Box 1029 Blindern, NO-0315 Oslo, Norway}

\author{Bjørn Anders Pettersson Reif}
\email{b.a.p.reif@its.uio.no}
\affiliation{Department of Technology Systems, University of Oslo, P.O. Box 70 Kjeller, NO-2027 Kjeller, Norway}
	
\begin{abstract}
This study investigates the Reynolds-number dependence of shock-induced flow through particle layers at 10\% volume fraction, using ensemble-averaged results from particle-resolved large eddy simulations. The advantage of using large eddy simulations to study this problem is that they capture the strong velocity shears and flow separation caused by the no-slip condition at the particle surfaces. The shock particle cloud interaction produces a reflected shock wave, whose strength increases with decreasing particle Reynolds number. This results in important changes to the flow field that enters the particle cloud. The results show an approximate proportionality between the mean flow velocity and the flow fluctuation magnitudes. Maximum particle drag forces are in excellent agreement with previous inviscid studies, and we complement these results with statistics of time-averaged particle forces as well as the variation of temporal oscillations. The results of this work provides a basis for development of improved simplified dispersed flow models.
\end{abstract}	
	
\maketitle

\section{Introduction}	

Interactions between shock waves and dense suspensions of particles occur in many engineering applications and industrial processes. Some examples are liquid and solid fuel engines \citep{chang2003,davis2013,ren2018}, heterogeneous explosives \citep{zhang2001}, blast mitigation \citep{chaudhuri2013,milne2014} and dust-explosion safety measures \citep{sichel1995}. Natural occurrences of shock-wave particle cloud interactions include volcanic eruptions \citep{chojnicki2006} and supernovae \citep{inoue2009,silvia2012}. 

The process of shock wave particle cloud interaction has recently received attention in both experimental \citep{wagner2012,theofanous2016,demauro2017} and numerical studies \citep{ling2012,houim2016,theofanous2017,shallcross2018,sugiyama2018}. Numerical studies have primarily utilized two different approaches. The first type employs Eulerian-Eulerian or Eulerian-Lagrangian methods to simulate flows with a large number of particles. The model equations are formulated to account for the volume and density of the dispersed phase. There are numerous issues with such simplified dispersed flow models, such as non-hyperbolic equation sets, as discussed in \citet{lhuillier2013,theofanous2017}. The second approach circumvents these problems by utilizing particle-resolved simulations. This is the method utilized in this study. Particle-resolved simulations are valuable because they provide highly resolved data, in both space and time, which enables a comprehensive analysis. The primary limitation of such simulations is that only small-scale problems are computationally feasible. 

There is a considerable body of work investigating shock wave particle cloud interaction using particle-resolved inviscid simulations \citep{regele2014,theofanous2018,mehta2018,mehta2019b,mehta2019,sen2018}. The number of studies that have included viscous effects in such simulations is, however, limited. Particle-resolved studies that have included viscous effects are, e.g., the two-dimensional simulations of \citet{naiman2007,hosseinzadeh2018} and the three-dimensional simulations of \citet{vartdal2018,osnes2019}. Viscous effects have not been analyzed directly in experimental studies either. There is, however, a moderate range of particle Reynolds numbers $\Rep$ where experimental results exist. Based on the shock wave strength, the particle Reynolds numbers in \citet{wagner2012} are approximately $2650-4200$. The study of \citet{demauro2017} spans $\Rep\approx900-5700$, whereas \citet{theofanous2016} investigated configurations in the range $\Rep\approx7000-70000$. Since these studies vary Reynolds number simultaneously with Mach number and ratio of layer thickness to particle diameter, isolating the Reynolds-number effects is difficult. 

The Reynolds number is a measure of the relative importance of inertial to viscous forces within the flow. Viscosity has an effect on the flow field even at very high particle Reynolds numbers, as long as the flow remains in the continuum regime. In particular, viscous effects are responsible for flow separation behind particles. This is in fact an important phenomenon in shock wave particle cloud interaction, as will be shown in this study. 

This study considers particle Reynolds numbers in the range $500-10000$. In incompressible, single-particle flows, this particle Reynolds number range spans three different flow regimes, by their definitions in \citet{tiwari2019}. These flow regimes are commonly labeled the vortex shedding regime ($270 \leq \Rep \leq 800$), the separating vortex regime ($800 < \Rep \leq 3000$), and the sub-critical regime ($3000 < \Rep < 3.4\times 10^5$). \citet{sakamoto1990} discusses the flow properties over this range and draws the following conclusions. In the vortex shedding regime, laminar vortices are  shed periodically. Above $\Rep=800$, vortex tubes and vortex loops are formed in the vortex sheet, and can interact with the large scale laminar vortices, introducing irregularity in the vortex shedding. Above $\Rep=3000$, the vortex sheet becomes turbulent, and the turbulence intensity increases with $\Rep$. As a result of this, the Strouhal number, i.e., the non-dimensional vortex shedding frequency, decreases with Reynolds number. At $\Rep=6000$ and above, the vortex sheet is fully turbulent and the Strouhal number is approximately constant. 

Compressible flows over isolated particles in this Reynolds number range have not received nearly as much attention. \citet{nagata2016,nagata2018} investigated $50\leq \Rep \leq 1000$. Notably, the drag coefficient increases by more than a factor of two when the flow goes from subsonic to supersonic in this $\Rep$ range. The Mach number, $\M$, also has a stabilizing effect on the flow, and \citet{nagata2018} found that at $\M=2$ the flow is stable to at least $\Rep=1000$.

It is not straightforward to apply these findings to shock-wave particle cloud interaction. Importantly, the incident flow becomes subsonic due to the generation of a planar, upstream propagating, reflected shock \citep{boiko1997,wagner2012,demauro2017}. The strength of this shock determines which flow conditions the particles are exposed to, and these flow conditions define a particle Reynolds number and a local Mach number that is not known \textit{a priori}. These parameters are likely more appropriate for classification of the flow through the particle cloud after the initial shock-induced transient than the ones based on flow properties behind the incident shock. Additionally, the flow Mach number increases with downstream distance within the particle cloud \citep{regele2014,theofanous2018,osnes2019} in a manner similar to that observed in Fanno flows \citep{emanuel1968}. The presence of nearby particles also affects the flow around each particle by changing the direction of the incoming flow, exposing it to particle wakes, etc.  

As will be shown in this study, the particle Reynolds number affects the level of particle-scale fluctuations within the particle cloud. These fluctuations are defined here as the deviations from the volume averaged velocity, and need not be turbulent fluctuations; a laminar particle wake generates significant fluctuations. When the particle concentration increases, the impact of the wake-induced fluctuations also increases. In particular, particle wakes start to comprise a significant part of the volume within the cloud. This causes a problem for models based on volume averaging. The volume averaged velocity is shifted towards the average particle velocity, and as a result, it represents neither the flow in immediate proximity of each particle, nor the freer flow between them. In the current work, we explore what this means for the fluctuation levels and particle drag coefficients. 

The current study is concerned with how the particle Reynolds number affects shock-wave particle cloud interaction. To this end, we perform particle resolved, three-dimensional, viscous simulations of the passage of a shock wave through a particle layer with volume fraction 0.1. The particle positions are assumed to be fixed over the time frame considered. We examine shock wave attenuation and reflection, mean flow fields and fluctuation statistics. We also explore the Reynolds number dependence of forces acting on the fixed particles, including both maximal values, time-averages and distributions of the drag coefficients. The results from the current study are important as a baseline for development of closures for simplified dispersed flow models, as well as the formulation of particle force models for shock-wave particle cloud interaction.  

This paper is structured as follows. In \cref{sec:goveq} we describe the equations governing the flow through the particle cloud, the mathematical framework used in the flow analysis and the computational method. \Cref{sec:setupandgrid} specifies the problem under investigation. The simulation results are presented and discussed in \cref{sec:results}, and \cref{sec:concludingremarks} contains concluding remarks.

\section{Governing equations\label{sec:goveq} and computational approach}
The motion of the fluid is governed by the compressible Navier-Stokes equations, which can be written
\begin{equation}
\pdts{\rho}+\pdk{\rho u_k}=0,
\label{eq:mass}
\end{equation}
\begin{equation}								
\pdt{\rho u_i}+\pdk{\rho u_i u_k}=-\pdis{p} + \pdjs{\sigma_{ij}},
\label{eq:momentum}
\end{equation}
\begin{equation}
\pdt{\rho E}+\pdk{\rho E u_k + p u_k}=\pdj{\sigma_{ij}u_i} - \pdk{\lambda\pdks{T}},
\label{eq:energy}
\end{equation}
where $\rho(\mathbf{x},t)$ is the mass density, in which $\mathbf{x}$ is the spatial coordinate vector and $t$ denotes time, $u(\mathbf{x},t)$ is the velocity, $p(\mathbf{x},t)$ is the pressure, $\sigma_{ij}(\mathbf{x},t)=\mu (\pdjs{u_i}+\pdis{u_j}-2\pdks{u_k}\delta_{ij}/3)$ is the viscous stress tensor,  $\mu(\mathbf{x},t)$ is the dynamic viscosity, $E(\mathbf{x},t)=\rho e + 0.5\rho u_ku_k$ is the total energy per unit volume, $e(\mathbf{x},t)$ is the internal energy per unit mass, $\lambda(\mathbf{x},t)$ is the thermal conductivity, and $T(\mathbf{x},t)$ is the temperature. Here, $\partial_t$ denotes partial differentiation with respect to time and $\partial_{i,j,k}$ denotes partial differentiation with respect to space. The subscripts denote tensor components, and repeated indices imply summation from 1-3. Internal energy, pressure and density are related by the ideal gas equation of state, with $\gamma=1.4$. Temperature and internal energy are related by a constant specific heat capacity. We assume that the viscosity varies with temperature as $\mu(\mathbf{x},t)=\mu^0(T/T^0)^{0.76}$, where $\mu^0$ is the viscosity at $T=T^0$. We further relate the thermal diffusivity to the viscosity through a constant Prandtl number of $0.7$. 

The analysis in this work is based on the volume averaged momentum conservation equations. A discussion of volume averaging for flows containing particles can be found in, e.g., \citet{crowe2011,schwarzkopf2015}. Since the configuration is statistically one-dimensional in space, only the streamwise momentum equation is relevant. It can be expressed
\begin{equation}
\begin{split}
\pdt{\alpha \phrho\fu_1} + \pdx{\alpha\phrho\fu_1\fu_1 + \alpha \php}=\pdx{\alpha\phavg{\sigma}_{11}}&\\-\pdx{\alpha\phrho\favg{R}_{11}} +\frac{1}{V}\int_Sp n_1 dS - \frac{1}{V}\int_S \sigma_{1k}n_kdS&, 
\end{split}
\label{eq:vavgmom}
\end{equation}
where $\alpha$ is the gas volume fraction, $\R_{ij}(\mathbf{x},t)=\wfavg{u_i''u_j''}$ is the velocity fluctuation correlation tensor, $V$ denotes the averaging volume, $n$ is the normal vector pointing into the continuous phase, and $S$ denotes the continuous phase boundary. Here, $\phavg{\cdot}$ denotes a phase average, $\wfavg{\cdot}$ denotes a Favre average, and fluctuations from the Favre average are denoted $\cdot''$. Furthermore, we denote the fluctuating kinetic energy per unit volume, $\phrho\R_{ii}/2$, by $k$, and mean kinetic energy per unit volume by $K$. The particle volume fraction is denoted by $\alp$.

In the analysis of the computational results, the temporal evolution is discussed in terms of a time-scale related to the passage of the shock through the particle layer. This time-scale is 
\begin{equation}
\tauL=L/u_\mathrm{s},
\label{eq:timescales}
\end{equation}
where $u_\mathrm{s}$ is the speed of the incident shock wave and $L$ is the length of the particle cloud. Furthermore, we use $t_0$ to denote the time point when the shock wave is at $x=0$. 

The particle drag coefficient will be used in the discussion of particle forces. It is
\begin{equation}
C_\mathrm{D} = \frac{\int_{S_\mathrm{p}}(p n_1 -\sigma_{1k}n_k)dS}{0.5\phrho \favg{u}_1^2A_\mathrm{p}},
\end{equation}
where $S_\mathrm{p}$ is the particle surface and $A_\mathrm{p}$ is the projected area of the particle in the flow direction.

\subsection{Computational method\label{sec:compmethod}}
The present simulations are conducted using the compressible flow solver "CharLES" \citep{bres2017}, developed by Cascade Technologies. The compressible, filtered, Navier-Stokes equations are solved on an unstructured Voronoi-based grid \citep{bres2018}. CharLES is a finite volume large-eddy simulation code. It computes fluxes between control volumes based on an entropy-stable modified Lax-Friedrich flux. A more detailed discussion of the entropy stable scheme can be found in \citet{masquelet2017}. A third order explicit Runge-Kutta scheme is applied to advance the solution in time.

\section{Set-up and computational grid\label{sec:setupandgrid}}

\begin{figure*}
	\centerline{\includegraphics[]{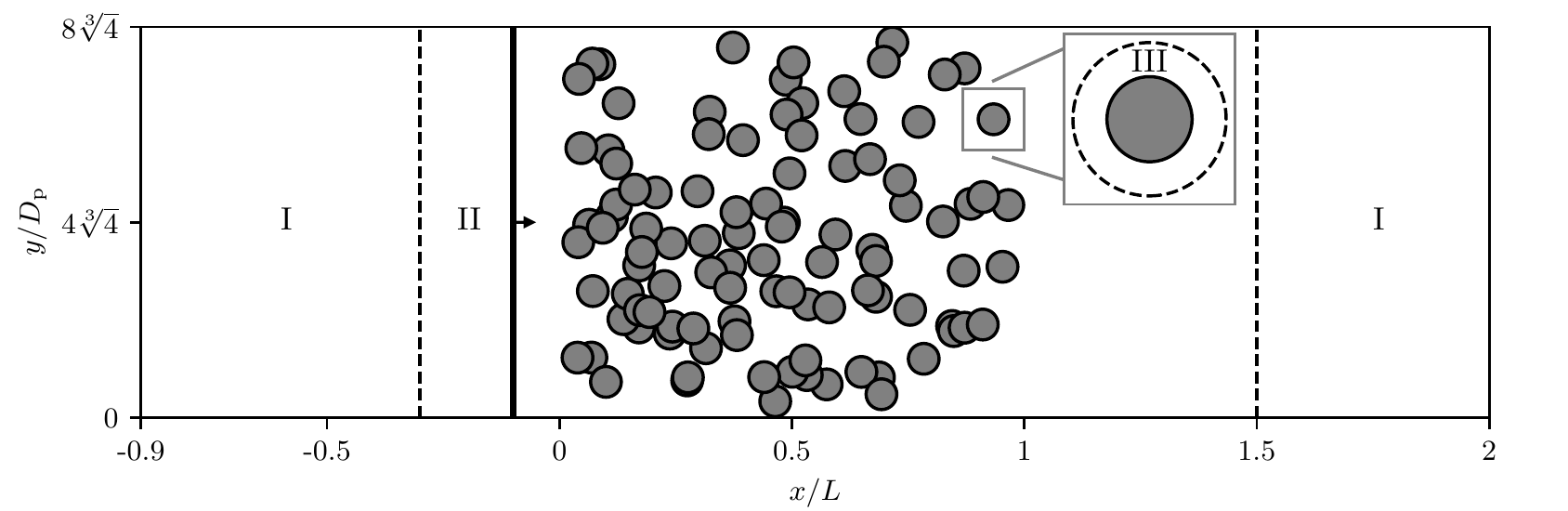}}
	\caption{Sketch of the computational set-up and domain. The particle layer is located between $0\leq x \leq L$, where the particle volume fraction is $\alp=0.1$. The thick solid line indicates the initial position of the shock wave, and the arrow shows its propagation direction. The roman numerals indicate the level of grid refinement, where the control volumes in region I are eight times larger than those in region II, and those in region II are again eight times larger than those in region III (by volume). The dashed lines indicate the boundaries between the refinement levels. Note that the vertical axis has been stretched for illustration purposes.}
	\label{fig:case_sketch}
\end{figure*}

We consider the interaction between a planar shock wave and a cloud of particles at different $\Rep$. \Cref{fig:case_sketch} shows a sketch of the computational set-up. The computational domain is $-0.9L \leq x \leq 2L$, $0 \leq y \leq 8\sqrt[3]{4}\Dp$ and $0 \leq z \leq 8\sqrt[3]{4}\Dp$, where $D_\mathrm{p}$ is the particle diameter. The particles are located within $0 \leq x \leq L$, and the particle volume fraction is $0.1$. In all configurations, the length of the particle layer and the particle diameter are related by $L=12\sqrt[3]{16}\Dp$. The particles are drawn randomly within this range, and we require that no particle boundary extends outside $0 \leq x \leq L$. The minimum distance between each particle is set to roughly $0.05\Dp$. The shock wave position is initially $-0.1L$, and is indicated by the thick line in the figure. Its direction of propagation is from left to right. 

The computational grid is an unstructured Voronoi-based grid, which is body-fitted to the particles. We use three different refinement levels, as indicated by the roman numerals in \cref{fig:case_sketch}. Refinement level I has control volumes eight times larger (by volume), than level II, and similarly for levels II and III. The length scale of the smallest control volumes (level III) is roughly $0.036\Dp$.  Due to the density of the particles, most of the region $0\leq x \leq L$ contains level III control volumes. The total number of control volumes is roughly $10^8$. The control volume sizes are similar to those used in \citet{osnes2019,osnes2019b}. Those studies investigated grid sensitivity based on the grid-convergence of the particle forces, resolution of viscous shear length scales, and convergence of spanwise velocity fluctuation correlations. 

This study investigates the effect of particle Reynolds number over the range $500-10000$ on the shock-induced flow through particle clouds. The particle Reynolds number is defined as
\begin{equation}
\Rep=\frac{\rho u\Dp}{\mu}.
\end{equation}
Throughout this paper, the Reynolds number is based on flow properties behind the incident shock wave, unless otherwise specified. Subscript IS will be used to denote post-shock values for the other flow variables.
The Reynolds number is systematically varied by altering the viscosity, rather than varying the particle diameter or the shock wave strength. This enables the use of the same computational grid for different Reynolds numbers. 

The initial conditions consist of two homogeneous spatial regions, which are separated by the shock wave at $x=-0.1L$. The pre-shock conditions are chosen to correspond to air at atmospheric conditions, i.e., $\rho^0=1.2048$ $\mathrm{kg/m^3}$, $u^0 = 0$ m/s, $p^0=1.01325\times10^5$ Pa. The viscosity $\mu^0$, is varied to obtain the desired particle Reynolds number. At $\Rep=5000$, this corresponds approximately to the viscosity of air at standard atmospheric conditions ($\mu^0 \approx 1.8\times10^{-5}\ \mathrm{kg/ms}$). The Mach number of the incident shock wave is $2.6$, and behind it is a homogeneous flow with the post-shock state, which is $\rho_\mathrm{IS}=4.16$ $\mathrm{kg/m^3}$, $u_\mathrm{IS}=633$ m/s, and $p_\mathrm{IS}=782$ kPa. 
At the $x=-0.9L$ boundary, we apply constant inflow conditions, corresponding to the post-shock state. At $x=2L$, we use a zero-gradient outlet. The spanwise boundaries are periodic. The spanwise domain lengths are both roughly 13 particle diameters, and this should be sufficient to hinder any periodic artifacts. The simulations are run for a total time of $3.75\tauL$.

Flow field statistics are obtained by averaging flow variables over predetermined volumes. In the current work, we utilize volumes spanning the computational domain in the $y$ and $z$ directions, with a streamwise length of $L/60\approx0.5\Dp$. 

A large number of particles is required in the spanwise directions to achieve well resolved statistics of the time-dependent flow field. Since only a limited number of particles is feasible to include in a single simulation, we perform an ensemble of simulations at each $\Rep$. The gas-phase variables are presented in terms of volume averaged quantities. The volume averages from the simulations are averaged over the ensemble of simulations, so that for example
\begin{equation}
\phrho(\mathbf{x},t)=\frac{1}{N_\mathrm{sim}}\sum_{i=0}^{N_{\mathrm{sim}}}\frac{1}{V^i(x)}\sum_{j=0}^{N_\mathrm{CV}^i(x)}\rho^j(t) V_\mathrm{CV}^j,
\end{equation}
where $N_\mathrm{sim}$ is the number of simulations (ten in this study), $N_\mathrm{CV}$ is the number of control volumes within the averaging volume $V(x)$ and superscripts denote quantities belonging to simulation $i$ or control volume $j$. Fluctuation correlations are computed implicitly as the difference between the products of the averages and the averages of the products. Thus, the velocity correlations are
\begin{equation}
\wfavg{u_i''u_j''}=\wfavg{u_iu_j}-\favg{u}_i\favg{u}_j.
\end{equation}
We consider only statistics up to second order moments, and we therefore use the convergence of the fluctuating kinetic energy, $k$, to estimate the required ensemble size. We find that the tenth simulation changes the statistics of $k$ by about $0.5\%$ on average. We therefore consider an ensemble of ten simulations at each $\Rep$ sufficient for the current analysis.

\section{Results\label{sec:results}}
This section presents the simulation results. \Cref{sec:shockwaveattenuation} examines the attenuation of the main shock wave and the strength of the reflected shock wave. \Cref{sec:flowfieldresults} discusses mean fields and fluctuations in the gas phase. Finally, \Cref{sec:particleforceresults} presents particle drag coefficient trends and distributions. 

\subsection{Shock-wave attenuation and reflection\label{sec:shockwaveattenuation}}

\begin{figure*}
	\centerline{
	\includegraphics[]{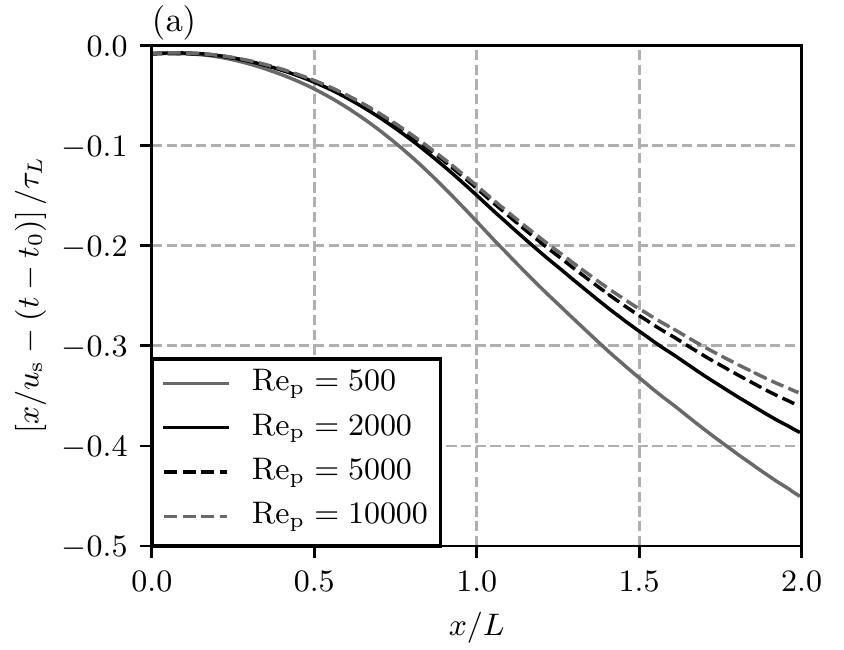}\includegraphics[]{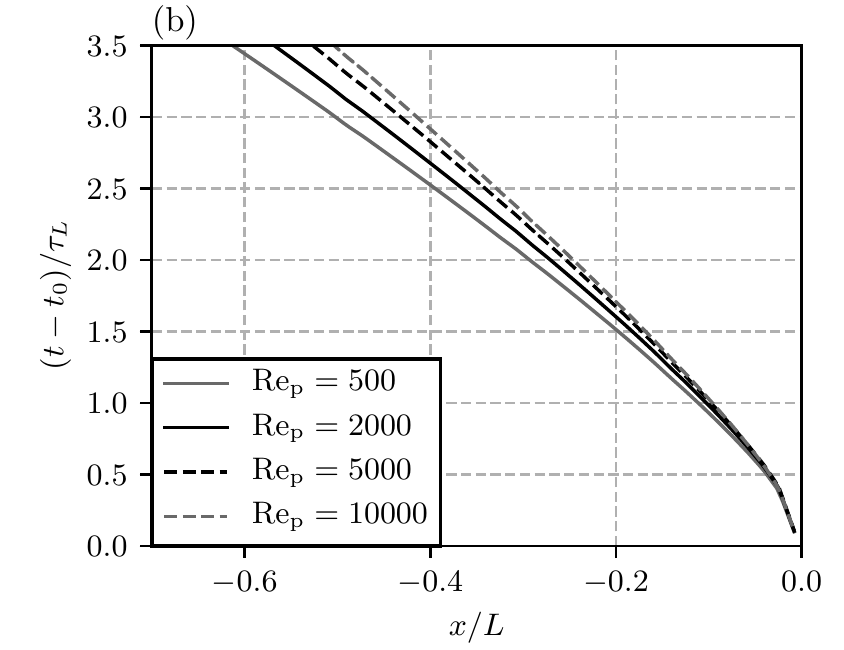}}
	\caption{(a) Main shock and (b) reflected shock position as a function of time for different $\Rep$. As opposed to the reflected shock, the variation in main shock position with $\Rep$ is small and is therefore plotted as the deviation from the trajectory of a hypothetical $\M=2.6$ shock wave with no attenuation.}
	\label{fig:shock_wave_positions}
\end{figure*}

\Cref{fig:shock_wave_positions} shows the main shock position as a function of time. The results are plotted as the deviation from the hypothetical trajectory of a shock wave with constant Mach number 2.6. This deviation is due to shock wave attenuation caused by the presence of the particle cloud. Both shock wave reflection and particle drag contributes to the attenuation. The arrival time at a given location is delayed for lower Reynolds numbers.

If there is to be a considerable effect of $\Rep$ on the shock wave attenuation, the impulse from viscous forces must be non-negligible on a time-scale comparable to the passage of the shock wave over a particle. Viscous forces are caused by the no-slip condition at the particle surface, which sets up a strong velocity gradient around the particle. An estimation of the importance of viscous forces on the shock wave attenuation can be obtained by comparing viscous and pressure forces acting on the particles during the time when the shock interacts with them. The average ratio of the viscous to pressure forces on the particles during the shock-particle interaction is given in \cref{tab:MrsMts}. It can be seen that for the lower Reynolds numbers, the viscous forces are appreciable even during the initial phase. 

The Mach number of the transmitted shock wave, based on the average transmitted shock wave speed over $L \leq x \leq 2L$, is also given in \cref{tab:MrsMts}. The Mach number is reduced by $5.3\%$ as $\Rep$ decreases from 10000 to 500. For the particle layer lengths examined here, the additional shock wave attenuation by viscous effects is only moderate, but it might become significant if longer layers are considered. 

\begin{table}
	\centering
	\caption{Average ratio of viscous to pressure forces acting on the particles (force ratio) over the time frame where the shock wave interacts with each particle, the Mach numbers of the transmitted ($\M_\mathrm{TS}$) and the reflected ($\M_\mathrm{RS}$) shock waves, the particle Reynolds number based on the flow state behind the reflected shock wave ($\mathrm{Re}_\mathrm{p,RS}$), and the local Mach number of the flow behind the reflected shock wave ($\M^*$).}
	\begin{tabular}{cccccc}
		\hline
		\hline
		$\Rep$ & Force ratio & $\M_\mathrm{TS}$ & $\M_\mathrm{RS}$ & $\mathrm{Re}_\mathrm{p,RS}$ & $\M^*$\\
		\hline
		$500$ & $0.142$ & $2.076$ & $1.580$ & $360$ & $0.379$ \\
		$2000$ & $0.059$ & $2.139$ & $1.558$ & $1495$ & $0.404$\\
		$5000$ & $0.028$ & $2.174$ & $1.530$ & $3904$ & $0.435$\\
		$10000$ & $0.015$ & $2.192$ & $1.516$ & $7967$ & $0.451$\\
		\hline
		\hline
	\end{tabular}
	\label{tab:MrsMts}
\end{table}

The position of the reflected shock wave as a function of time is also shown in \cref{fig:shock_wave_positions}. Since the incident flow is supersonic, the reflected shock wave moves slowly in this reference frame. Its strength increases with decreasing Reynolds number. This is expected, since the magnitude of the particle drag increases with reduced Reynolds number. There is a larger resistance for the flow to pass through the particle layer due to the viscous forces, which builds up pressure in the flow. Therefore the reflected shock strength must increase. Note that the strength of the reflected shock increases with time. 

The average Mach number of the reflected shock is given in \cref{tab:MrsMts}. It is based on the average speed of the reflected shock wave over the time-frame where it is located upstream of $x=-0.1L$. The lower Reynolds numbers have higher reflected shock Mach numbers, and the difference between $\Rep=10000$ and $\Rep=500$ is $4\%$. 

\Cref{tab:MrsMts} also shows the particle Reynolds number and the local Mach number of the flow based on the properties behind the reflected shock, assuming a constant reflected shock Mach number. For classification of the late-time flow field in the particle cloud, these two parameters are more appropriate than the Mach number behind the incident shock wave and $\Rep$ based on $\rho_\mathrm{IS}$, $u_\mathrm{IS}$ and $\mu_\mathrm{IS}$. The change in Reynolds number is moderate, and does not imply any qualitative changes in the flow, based on the characterization in single-particle studies. The change in local Mach number is more important because the flow behind the incident shock is supersonic while the flow behind the reflected shock is subsonic with a relatively low Mach number.

\subsection{Flow fields and fluctuations\label{sec:flowfieldresults}}

\begin{figure*}
	\centerline{\includegraphics[]{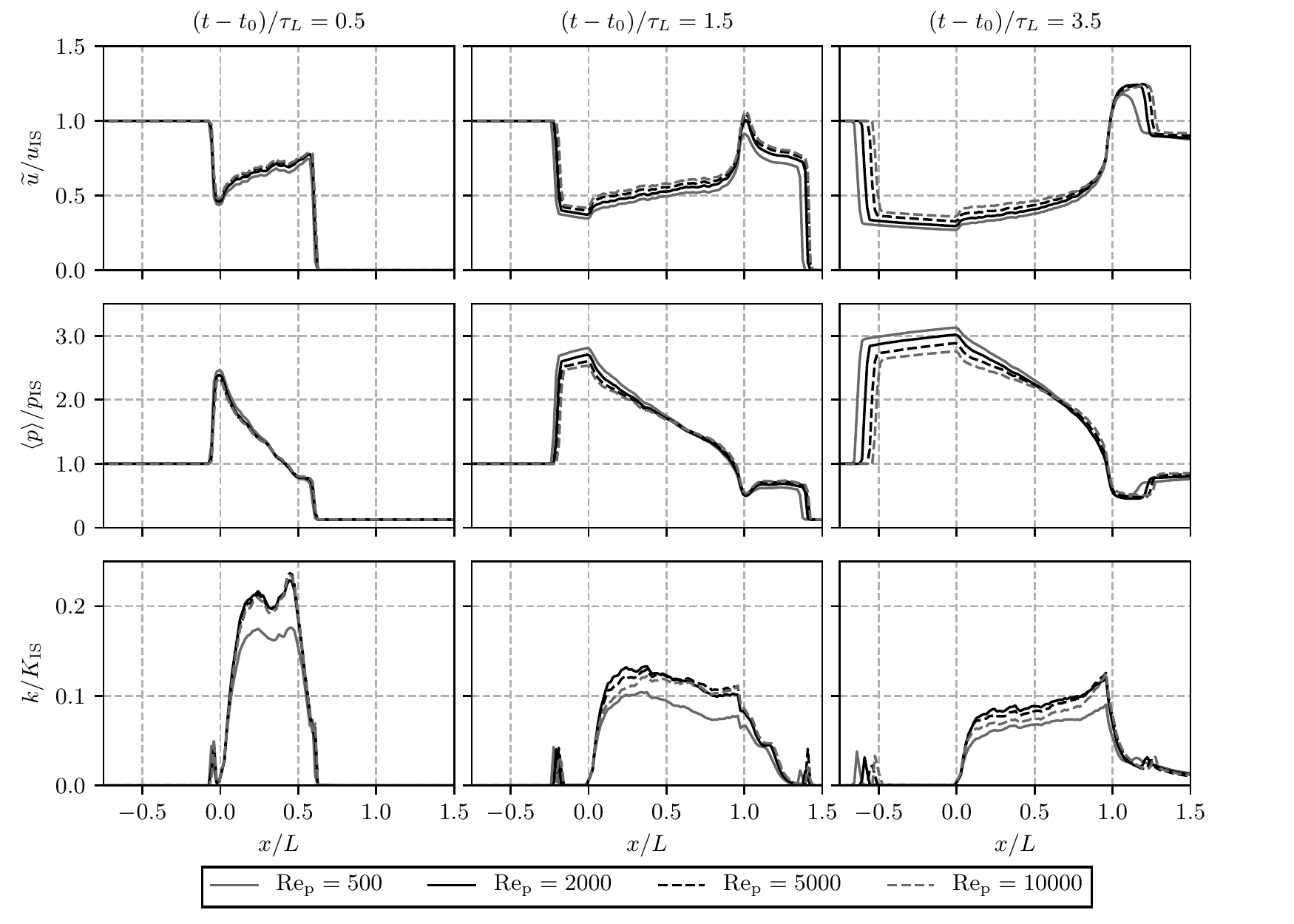}}
	\caption{Streamwise velocity, pressure and fluctuating kinetic energy at three times for different $\Rep$. The first column shows a time where the shock wave is still inside the particle layer, the second where it has recently exited the layer, and the final column is close to the end of the simulation.}
	\label{fig:flow_field_time_series}
\end{figure*}

\begin{figure*}
	\centerline{\includegraphics[]{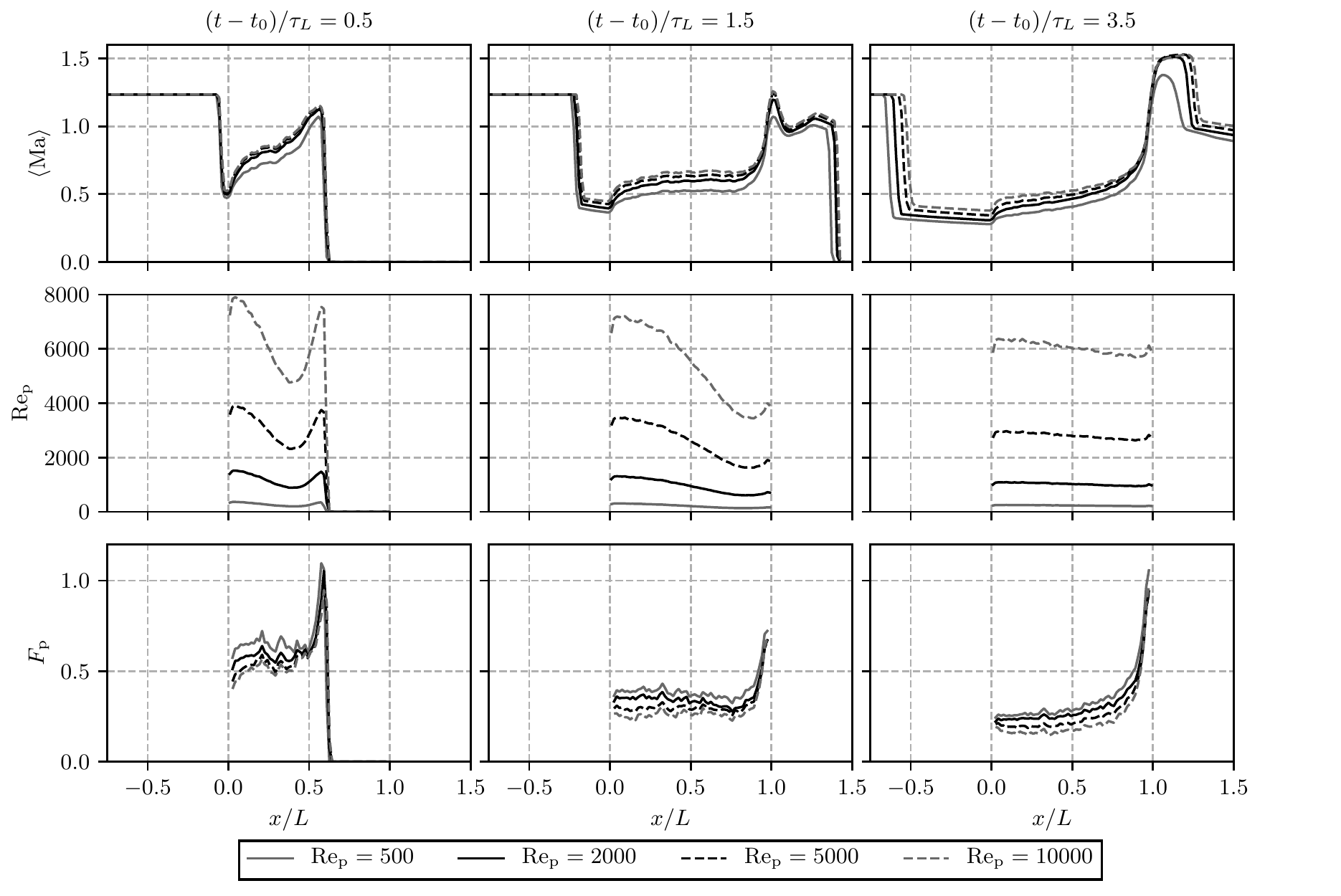}}
	\caption{Mach number, local particle Reynolds number, and average normalized particle forces at three time points for different $\Rep$. The first column shows a time point where the shock wave is still inside the particle layer, the second where it has recently exited the layer, and the final column is close to the end of the simulation. }
	\label{fig:flow_field_time_series_2}
\end{figure*}

\Cref{fig:flow_field_time_series} shows the normalized streamwise velocity, pressure and fluctuating kinetic energy at three different time points. The fluctuating kinetic energy is normalized by the inflow kinetic energy. Similarly, \cref{fig:flow_field_time_series_2} shows the Mach number, particle Reynolds number and normalized average streamwise particle forces, i.e.,
\begin{equation}
F_\mathrm{p}=\frac{1}{0.5N_\mathrm{p}(x)(\rho u^2)_\mathrm{IS}A_\mathrm{p}}\sum_{j=0}^{N_\mathrm{p}(x)}\int_{S_j}(p n_1 - \sigma_{1k}n_k)dS,
\end{equation}
where $N_\mathrm{p}(x)$ is the number of particles within the control volume centered at $x$, and $S_j$ is the surface of particle $j$. 

The fluctuating kinetic energy is the only flow quantity in \cref{fig:flow_field_time_series,fig:flow_field_time_series_2} that has a non-monotone dependence on $\Rep$. The highest level of fluctuating kinetic energy is found for $\Rep=2000$. The three highest Reynolds numbers are very similar at early time points, but the difference increases with time. 

The Mach number plots reveal that the flow becomes sonic just upstream of the downstream particle cloud edge. Similar observations have been made in other studies \citep{theofanous2017,theofanous2018,osnes2019}, and this behaviour is very similar to that found in Fanno flows. As discussed in \citet{emanuel1968}, the Mach number increases with downstream distance in Fanno flows due to the decreasing pressure. If the duct is sufficiently long, the flow chokes at the downstream exit, just like we observe here. 

In \cref{fig:flow_field_time_series_2} it can be seen that the particle forces increase drastically at the downstream layer edge for the two latest time points shown. The Reynolds number increases slightly as well, but is insufficient for explaining the increase in particle forces. Instead, the higher particle forces are due to the increased Mach number. 

\begin{figure}[h]
	\centering
	\includegraphics[]{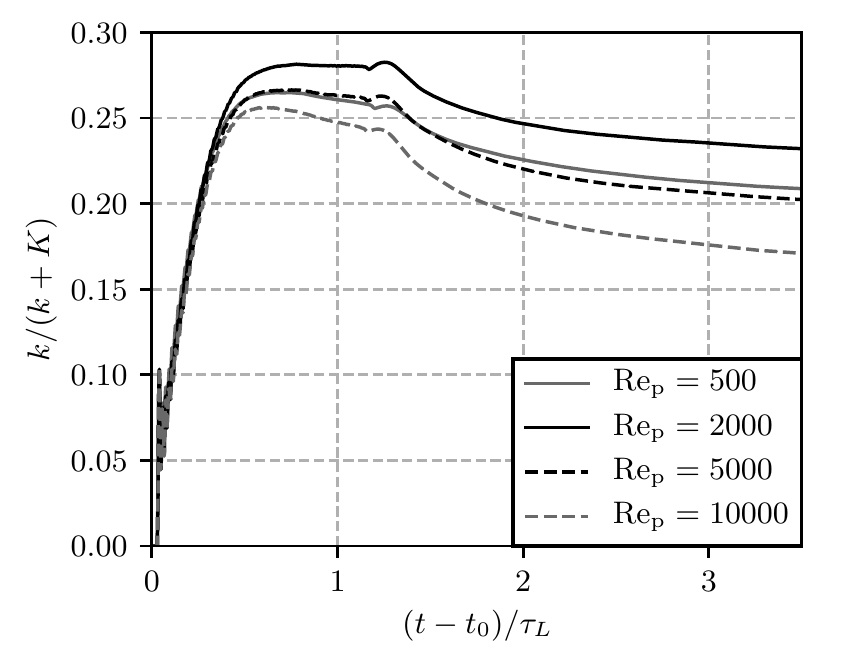}
	\caption{Ratio of fluctuating to total kinetic energy integrated over the particle layer, as a function of time, for different Reynolds numbers.}
	\label{fig:integral_k_div_K_interior_over_time}
\end{figure}

An important part of the high speed flow through particle clouds is the particle-scale fluctuations. These are commonly neglected in Eulerian-Lagrangian and Eulerian-Eulerian simulations of shock-wave particle cloud interactions, see , e.g., \citet{ling2012,houim2016,theofanous2017,sugiyama2018}. The gradient of the streamwise velocity fluctuations is an integral part of the momentum balance around the particle cloud edges. \citet{osnes2019} showed that at the upstream particle cloud edge, the streamwise Reynolds stress gradient is of equal importance as the particle forces after the initial shock-related transient. As can be seen in \cref{fig:flow_field_time_series}, the fluctuating kinetic energy has a very sharp gradient, and its magnitude is considerable, especially at early times. \Cref{fig:integral_k_div_K_interior_over_time} shows the ratio of fluctuating kinetic energy to total kinetic energy, integrated over the interior of the particle cloud, as a function of time. It can be seen that the ratio exceeds $0.25$ for all Reynolds numbers around $(t-t_0)/\tau_L=0.5$, i.e., when the shock wave is slightly less than halfway through the particle cloud. When the shock wave exits the cloud, the ratio decays slowly over time. The trend with $\Rep$ is not monotone, which is also the case for the fluctuating kinetic energy, c.f. \cref{fig:flow_field_time_series}. \citet{regele2014} showed that the pressure field computed in a one-dimensional Eulerian-Lagrangian simulation of shock-wave particle cloud interaction corresponded to the sum of the volume averaged pressure and the Reynolds stress in the their two-dimensional simulations. The overestimation of the pressure in their one-dimensional simulations is therefore the result of neglecting the energy deposited in velocity fluctuations. For the current simulations, the ratio $k/p$ takes values up to $0.25$ when the shock wave is inside the particle cloud. It remains non-negligible even at late times, where the value is between $0.02$ and $0.14$ within the particle cloud. The ratio is lower for $\Rep=500$ than the other Reynolds numbers, which are quite similar.

\begin{figure*}[h]
	\centerline{
		\includegraphics[]{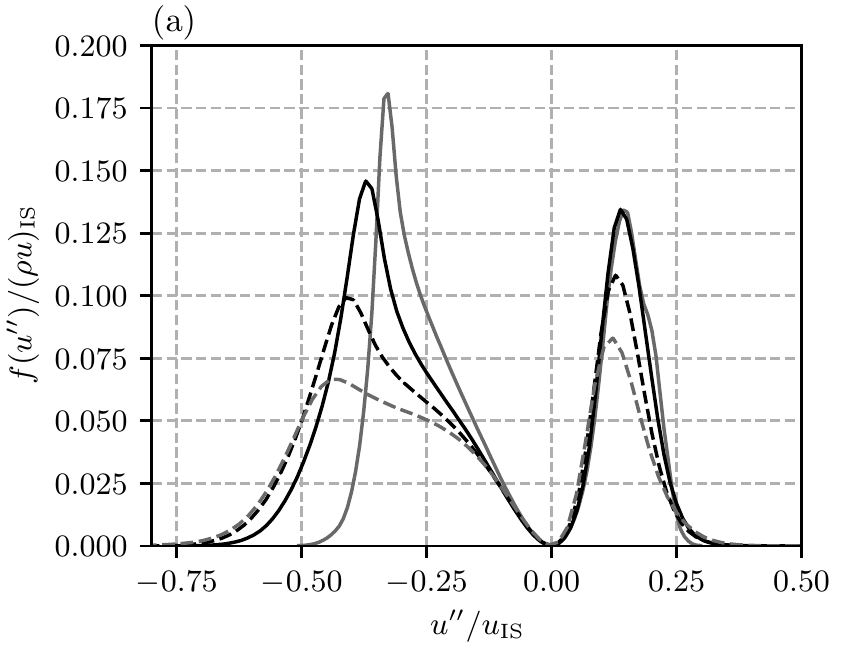}\includegraphics[]{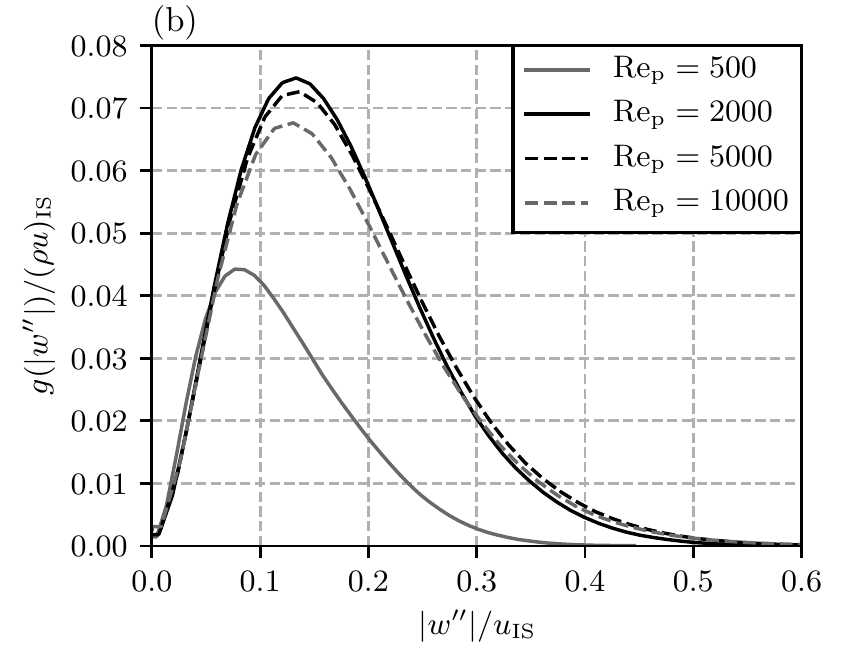}}
	\caption{Contribution to (a) the streamwise Reynolds stress and (b) the spanwise Reynolds stresses by different velocity fluctuations for different Reynolds numbers. Data for $0.25\leq x/L \leq 0.75$ at $t=3.75\tauL$.}
	\label{fig:Streamwise_Restress_contributions}
\end{figure*}

\begin{figure*}[h]
	\centerline{\includegraphics[]{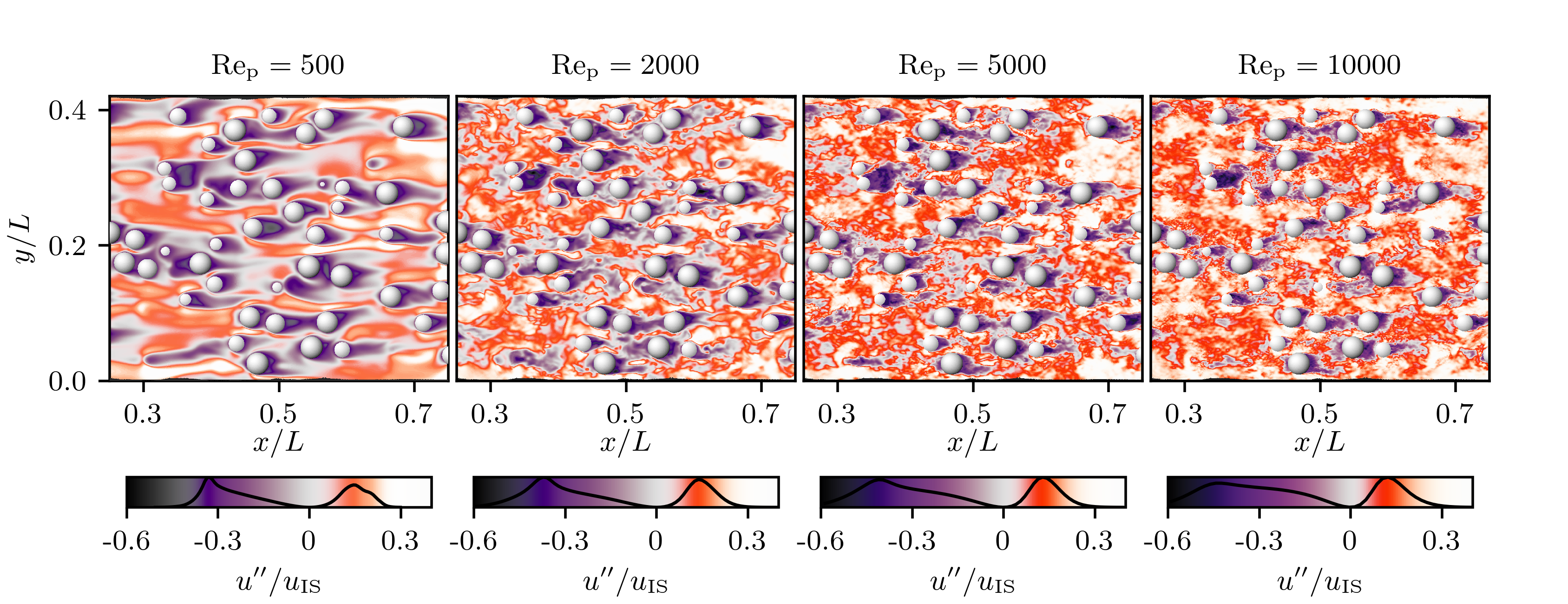}}
	\caption{Contours of $f$ at $t=3.75\tauL$ for one realization at each $\Rep$. The color indicates the velocity fluctuation and $f$ is used to set the color saturation, as indicated by the curve in each colorbar located below the plot. Thus areas with high color saturation have the highest contribution to the Reynolds stress. }
	\label{fig:Streamwise_Restress_contributions_contours}
\end{figure*}

Currently, there are very few models for flow field fluctuations that have been developed specifically for shock-wave particle cloud interaction. It is therefore useful to accurately identify flow phenomena that contribute to the fluctuation correlations, so that model development can be based on a fundamental understanding of the flow. In the current work, we consider only the velocity fluctuations.  
\Cref{fig:Streamwise_Restress_contributions} shows the contribution to the Reynolds stress by velocity fluctuations of different magnitudes at $t=3.75\tauL$. The figure shows the function $f$, which is defined by 
\begin{equation}
\phrho \R_{11}=\int_{-\infty}^{\infty}f(u'')du''.
\label{eq:f_definition}
\end{equation}
$f$ is a measure of the importance of velocity fluctuations of a given amplitude with regards to the velocity fluctuation correlations. For $\Rep=500$, the peak in $f$ is very sharp, and the contribution from  negative fluctuations below the peak is negligible. As the Reynolds number increases, the peak decreases. It moves towards more negative fluctuations because the mean flow velocity increases. The peak also broadens, and the contribution from strong negative fluctuations increases. The total contribution by negative fluctuations increases with Reynolds number, comprising $62.9\%$, $67.1\%$, $69.0\%$ and $69.2\%$ for Reynolds numbers $500$, $2000$, $5000$ and $10000$, respectively. The contribution from positive fluctuations can be seen to have a non-monotone dependence on Reynolds number over the range studied here. The peak value is very similar for the two lowest Reynolds numbers. When the Reynolds number is increased further, the peak value decreases to about $80\%$ and $65\%$ of the $\Rep=2000$ peak. Similar to the negative fluctuations, the curve broadens, but the effect is small. 

A similar plot of $f$ was used in \citet{osnes2019} to argue that the bulk contribution to the Reynolds stress was due to the circulation region behind each particle. Those results were for a flow configuration very similar to the $\Rep=5000$ case considered here. \Cref{fig:Streamwise_Restress_contributions_contours} shows contours of $f$ in space and confirms that the circulation regions are those with the highest contributions to the streamwise Reynolds stress. The flow in the circulation region has an average velocity of zero and therefore differs drastically from the mean flow speed. This implies that the volume of the circulation region is an important parameter for determining the streamwise Reynolds stress magnitude. It is also evident that most of the flow volume contains velocities that are higher than the volume averaged velocity. This effect arises because the separation regions shift the average velocity away from the velocity of the flow between the particles. At $\Rep\geq2000$ there are very few areas within the particle cloud that have a velocity equal to the volume-averaged velocity.

\begin{figure*}[h]
	\centerline{\includegraphics[]{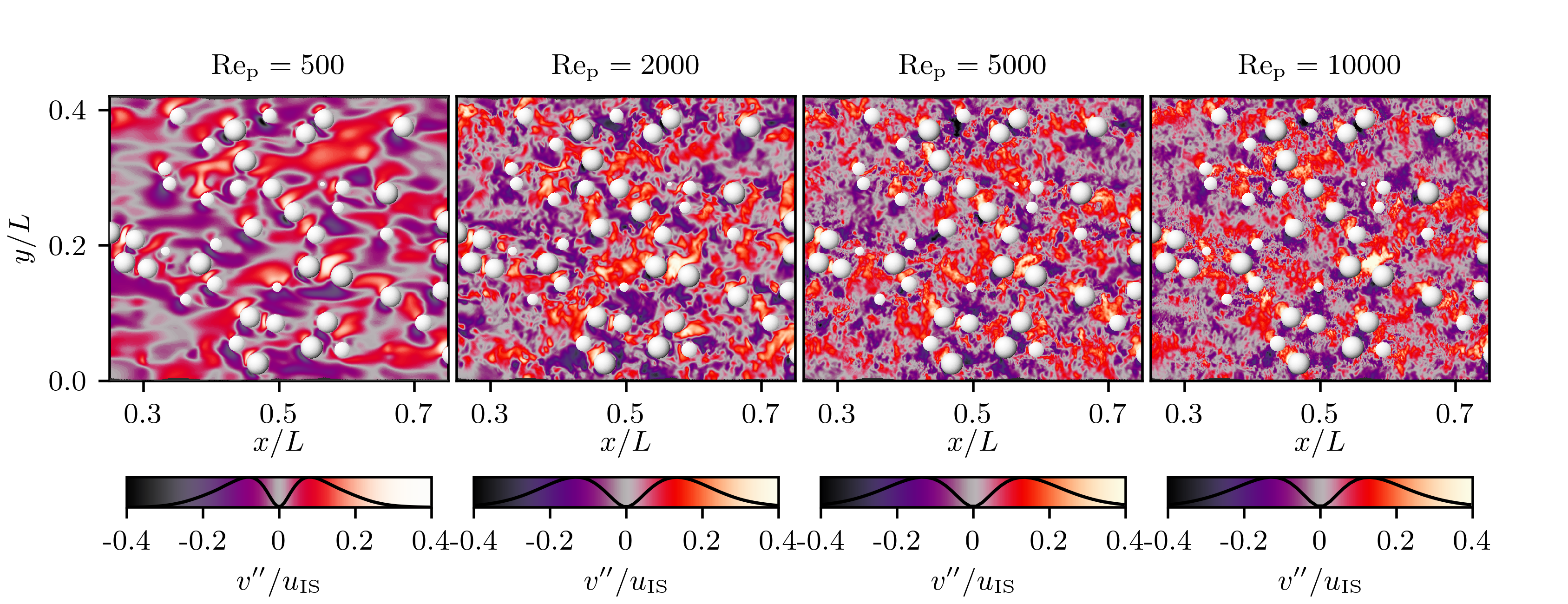}}
	\caption{Contours of $g$ at $t=3.75\tauL$ for one realization at each $\Rep$. The color indicates the velocity fluctuation and $g$ is used to set the color saturation, as indicated by the curve in each colorbar located below the plot. Thus areas with high color saturation have the highest contribution to the Reynolds stress.}
	\label{fig:Spanwise_Restress_contributions_contours}
\end{figure*}


\Cref{fig:Streamwise_Restress_contributions} also shows the contribution to the spanwise Reynolds stresses by spanwise velocity fluctuations of different magnitudes. The function $g(|w''|)$ is the analogue to $f(u'')$ for spanwise fluctuations, and is averaged over the two spanwise velocity components. Additionally, since $g(|w''|)$ should be symmetric around $w''=0$, we plot the contribution as a function of the fluctuation magnitude. The shapes of the curves are quite similar for all Reynolds numbers, but the magnitude is significantly lower for $\Rep=500$. The location of the peak value is at $0.08$ for $\Rep=500$ and around $0.135$ for the other Reynolds numbers. The trend with $\Rep$ is non-monotone, and the highest magnitude is obtained at $\Rep=2000$. This is similar to the trend for positive streamwise fluctuations, but the reduction at higher $\Rep$ is less significant for the spanwise fluctuations.

The areas that contribute most to the spanwise Reynolds stress components are shown in \cref{fig:Spanwise_Restress_contributions_contours}. It can be seen that at $\Rep=500$, the fluctuations are located in large coherent regions. These are significantly larger than a single particle, which indicates that it is the inter-particle spacing rather than the particle size that is important for the spanwise Reynolds stress components. The same phenomenon is seen for the higher Reynolds numbers, but to a lesser degree. At $\Rep=10000$, the scale of the coherent regions only appears slightly larger than a single particle. The regions contributing to the spanwise fluctuations are not the same as those that are important for the streamwise fluctuations. The streamwise fluctuations are primarily particle wakes, whereas the spanwise fluctuations are mostly due to flow deflection around particles and streaming through channels between them. The spanwise fluctuations are therefore located slightly in front and to the side of each particle, as well as in the open spaces within the cloud. 

\begin{figure*}
	\centerline{
		\includegraphics[]{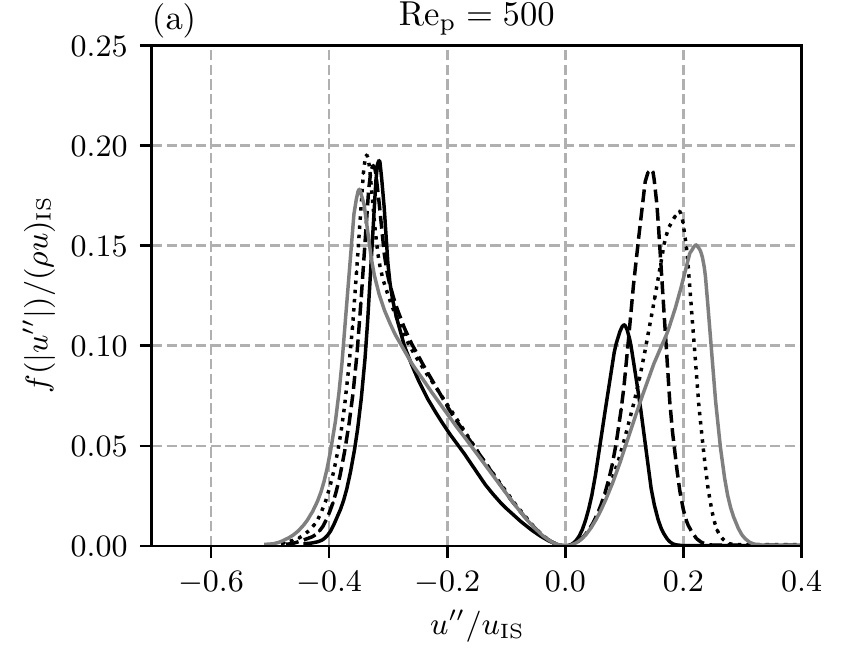}\includegraphics[]{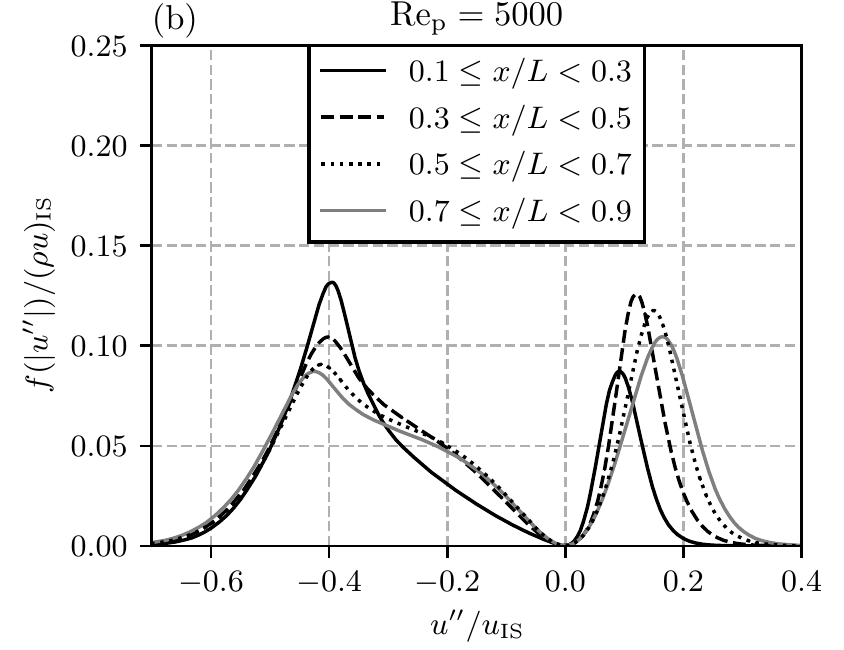}}
	\caption{Contribution to the streamwise Reynolds stress by different velocity fluctuations at $t=3.75\tauL$ for different streamwise positions. (a) $\Rep=500$. (b) $\Rep=5000$.}
	\label{fig:Restress_contributions_positions}
\end{figure*}

\begin{figure*}
	\centerline{
		\includegraphics[]{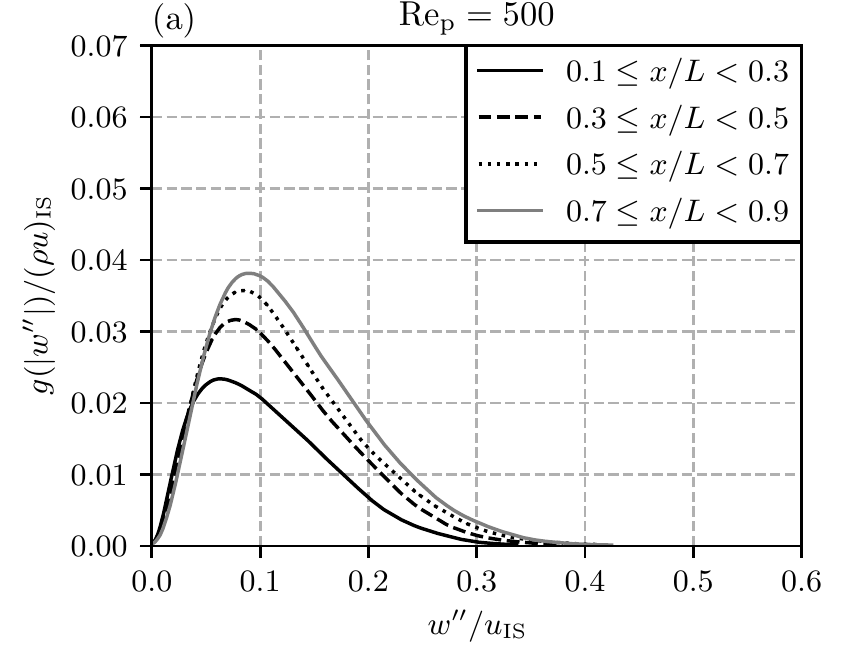}\includegraphics[]{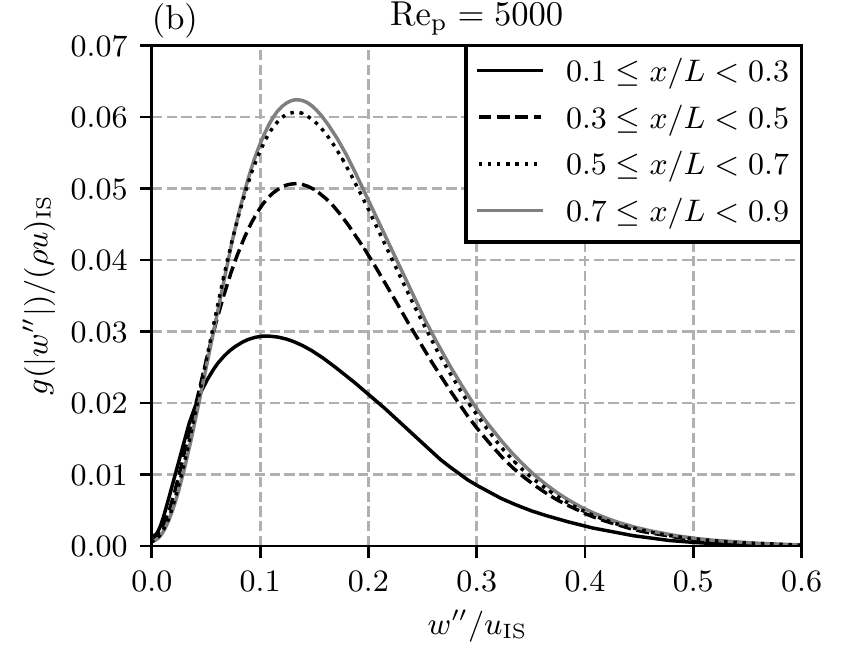}}
	\caption{Contribution to the spanwise Reynolds stress by different velocity fluctuations at $t=3.75\tauL$ for different streamwise positions. (a) $\Rep=500$. (b) $\Rep=5000$.}
	\label{fig:Spanwise_Restress_contributions_positions}
\end{figure*}

In \Cref{fig:Restress_contributions_positions} and \Cref{fig:Spanwise_Restress_contributions_positions}, the streamwise and spanwise Reynolds stress contributions are plotted at different streamwise locations within the particle layer. $\Rep=2000$ and $10000$ have very similar trends as $\Rep=5000$ and are therefore not shown. It can be seen that negative streamwise fluctuations at $\Rep=500$ have very little variation with streamwise position. On the other hand, positive fluctuations shift towards higher magnitudes as distance increases, and the curve broadens. This is also the case for $\Rep=5000$, although the shifts are slightly lower. The shift in positive direction is related to the flow acceleration, which is stronger for $\Rep=500$ than for $\Rep=5000$. For $\Rep=5000$, the shape of the curve for negative fluctuations changes with downstream distance. Smaller magnitudes become more important, presumably due to increasingly turbulent flow. 

For spanwise fluctuations, we find that at $\Rep=500$, the magnitude increases with downstream distance throughout the particle cloud. Additionally, the peak location shifts towards higher fluctuation magnitudes. At $\Rep=5000$, the trend is similar up to $x/L=0.5$, but further downstream distance has little effect on the fluctuations.

\begin{figure*}
	\centerline{
	\includegraphics[]{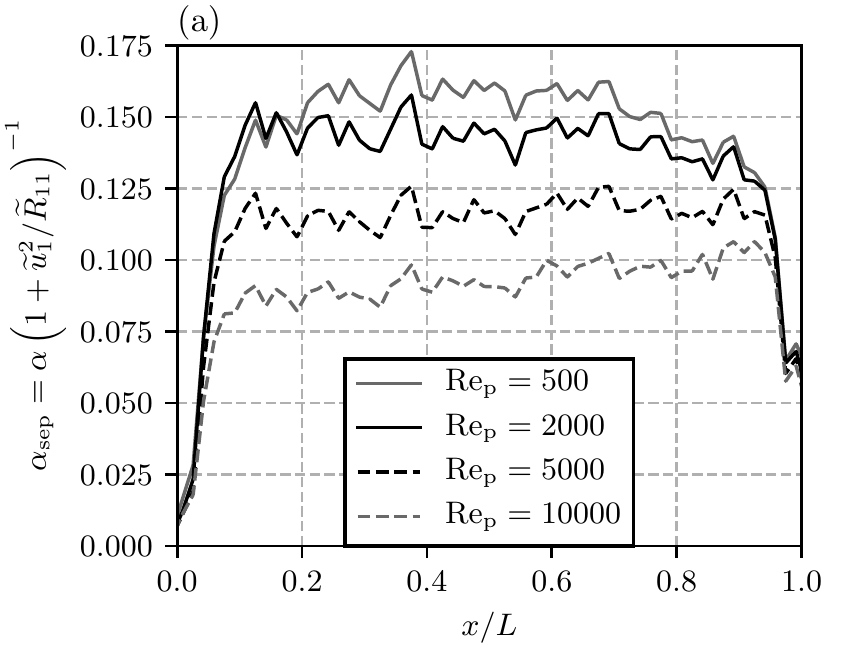}\includegraphics[]{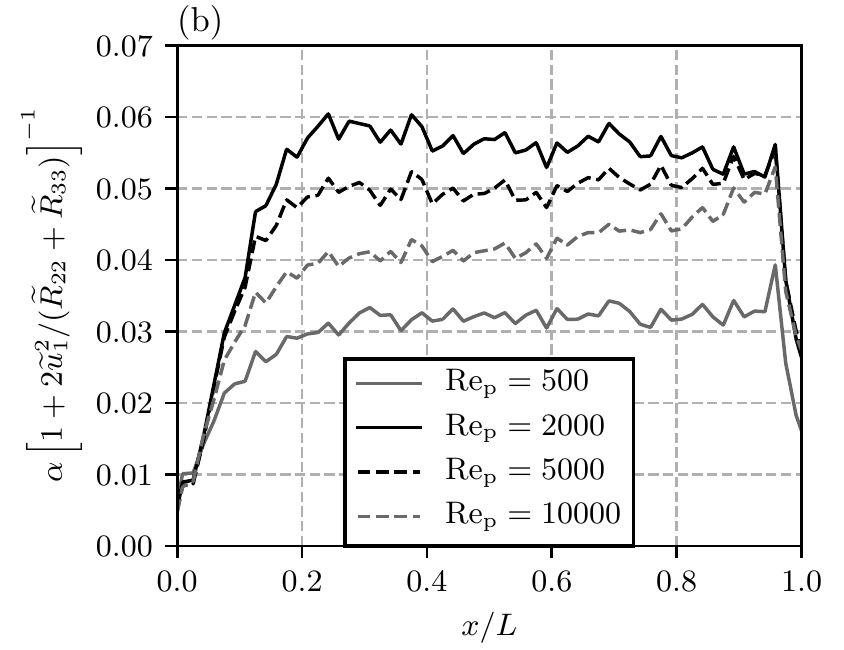}}
	\caption{(a) $\alpha_\mathrm{sep}$ and (b) the analogous function for spanwise fluctiations for different $\Rep$, at $t=3.75\tauL$.}
	\label{fig:separationvolume}
\end{figure*}

\citet{osnes2019} found that the streamwise fluctuations in the interior of the particle cloud could be represented by
\begin{equation}
\phavg{u_1''u_1''} = \favg{u}_1^2\frac{\alps}{\alpha-\alps},
\label{eq:alphasep}
\end{equation}
where $\alps$ is an estimate of the average volume of the circulation region behind each particle. This model approximates the flow as two constant states. One is the flow between the particles and the other is the separated flow behind a particle, where the velocity is assumed to be zero. We note that this model does not predict zero velocity fluctuations when $\alpha$ goes to one, nor is it Galilean invariant, and therefore needs a modification if it is to be used in simplified dispersed flow models. A way to correct this is to replace $\favg{u}_1$ by the relative velocity between the particles and the gas, and replace $\alps$ by $\alpha_\mathrm{p}C$  in \cref{eq:alphasep}, where $C$ is a constant or a function of dimensionless flow variables. \Cref{fig:separationvolume} shows $\alps$ at the end of each simulation, estimated from the simulation results using \cref{eq:alphasep}. $\alps$ changes rapidly close to the cloud edges, but over the interior of the cloud it has an approximately constant value. $\Rep=500$ deviates slightly more from a constant value than the higher Reynolds numbers. The trend with $\Rep$ is monotone over this range, and lower Reynolds numbers have larger separation volumes. The trends found here agree qualitatively with observations from single sphere studies in the incompressible regime, where the separation length, i.e., the average distance from the particle to the end of the separated flow region, was found to be roughly 30\% shorter at $\Rep=10000$ than at $\Rep=3700$ \citep{rodriguez2013}. Indications of a similar trend were also found for compressible flows, where $\Rep=1000$ at $\M=0.8$ and $1.2$ had a shorter separation length than $\Rep=750$ \citep{nagata2018}. 

\Cref{fig:separationvolume} also shows the analogous function to $\alps$ for the spanwise fluctuations. It can be seen that this function has little variation over $0.2\leq x/l \leq 0.95$. It does not follow from the derivation of \cref{eq:alphasep} that the spanwise fluctuations should vary this way. The results should be interpreted as confirming that the velocity fluctuations vary proportionally to the mean streamwise flow velocity throughout most of the particle layer. We note that for the spanwise fluctuations, there is a longer distance at the start of the layer where the function increases with distance. We suspect that this is because spanwise fluctuations are more related to inter-particle distance than particle diameter. If this is the case, it is not surprising that the build-up distance is longer, since the inter-particle distance is larger than the particle diameter on average. In contrast to the monotonous behavior of $\alps$ with $\Rep$, there is a non-monotone trend for the spanwise fluctuations. We also find that for $\Rep=10000$, the function begins to increase towards the end of the layer. We suspect that this is the result of increasingly turbulent flow. 

The simple relationship between the mean flow velocity and the velocity fluctuations is surprisingly robust across different flow conditions. The flow field within the particle cloud features rapidly varying velocity-, pressure- and density-magnitudes as well as varying gradients. Additionally, the Mach numbers vary from 0.3 to 1. The insensitivity to these variations is encouraging for development of velocity fluctuation correlation models that can be used as closures for simplified dispersed flow models.   

\subsection{Particle drag\label{sec:particleforceresults}}
\begin{figure*}
\centerline{
\includegraphics{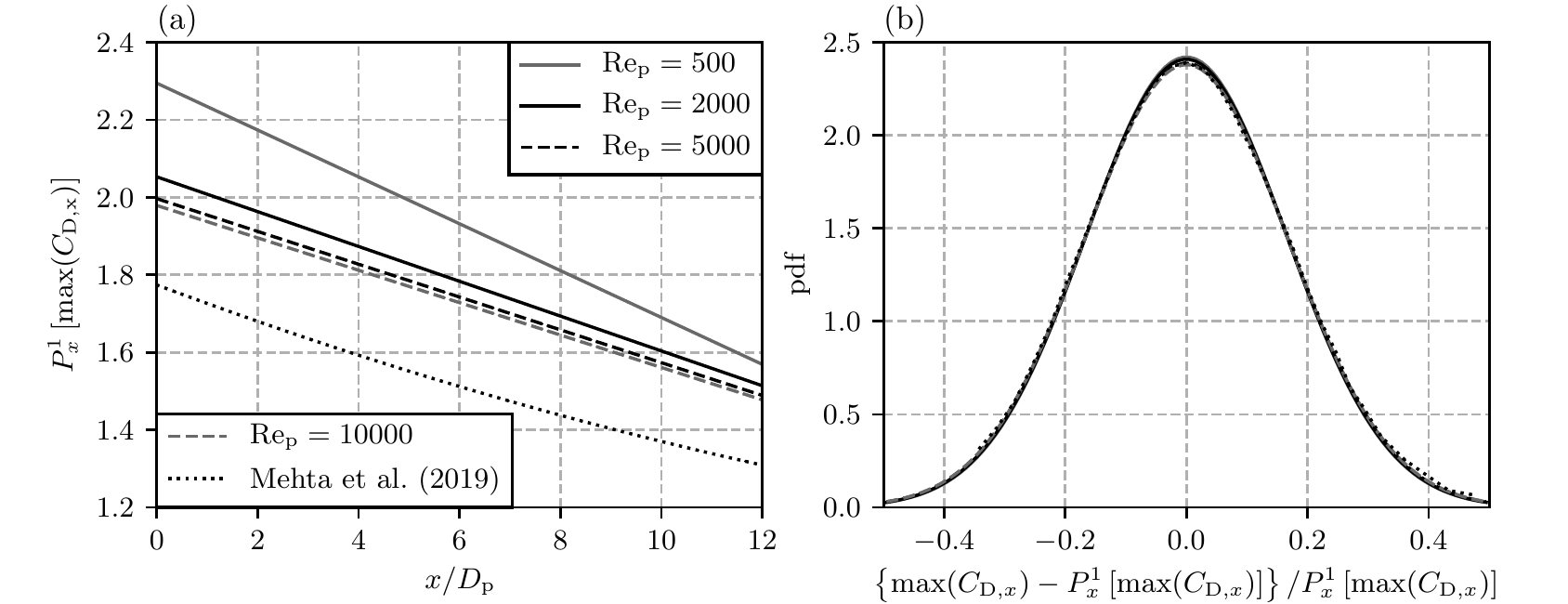}}
\caption{(a) Least squares first order polynomial fit of peak streamwise drag-coefficients, denoted $P_x^1\left[\max(\CDx)\right]$, as a function of $x$ for different Reynolds numbers. (b) Best normal distribution fit to the deviation of the peak drag coefficients from $P_x^1\left[\max(\CDx)\right]$. The plots also show the results obtained in inviscid $\M=3$ simulations by \citet{mehta2019}. In order to compare against those results, only particles between $0\leq x/\Dp\leq 12.5$ have been included for the simulations in this work. Note that for the results shown in this figure, drag coefficients are based on $\rho_\mathrm{IS}$ and $u_\mathrm{IS}$}
\label{fig:peak_cdx_distribution}
\end{figure*}

\begin{figure*}
	\centerline{
	\includegraphics[]{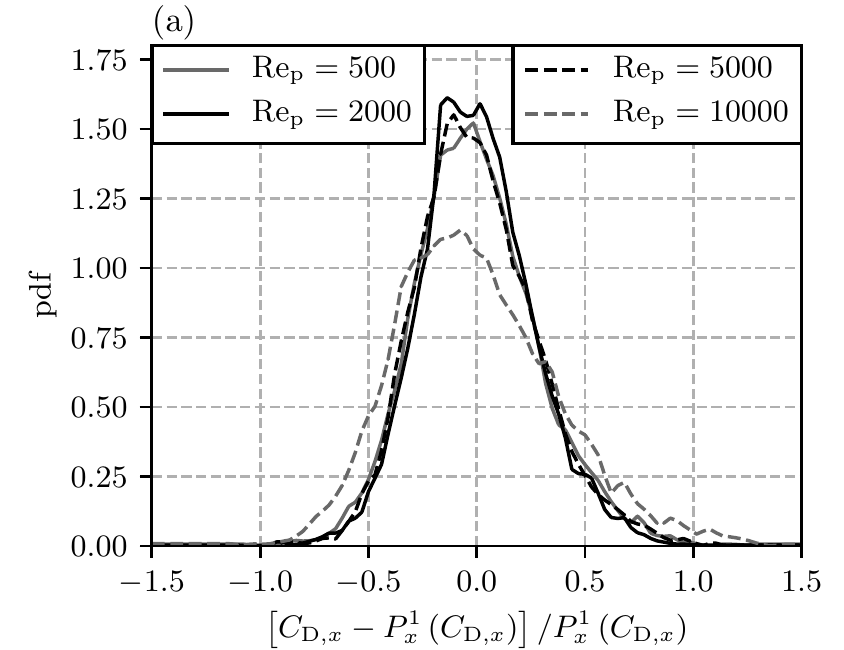}\includegraphics[]{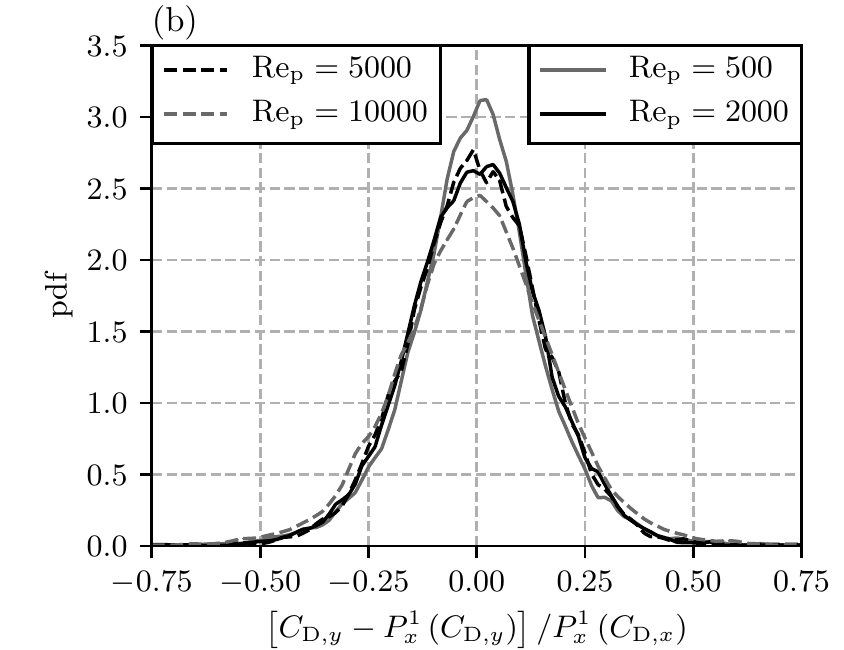}}
	\caption{Distribution of the time-averaged drag coefficients for $t>1.875\tauL$. (a) Streamwise drag coefficient. (b) Spanwise drag coefficient. The linear trend with $x$ has been subtracted. Note the different scale in the plots.}
	\label{fig:cdx-mean-distribution}
\end{figure*}

\begin{figure*}
	\centerline{
	\includegraphics[]{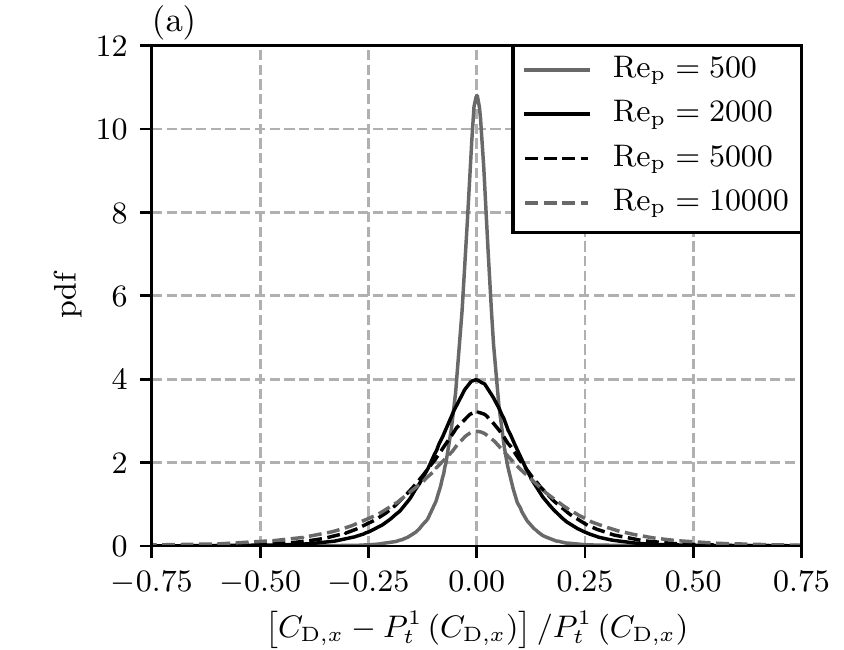}\includegraphics[]{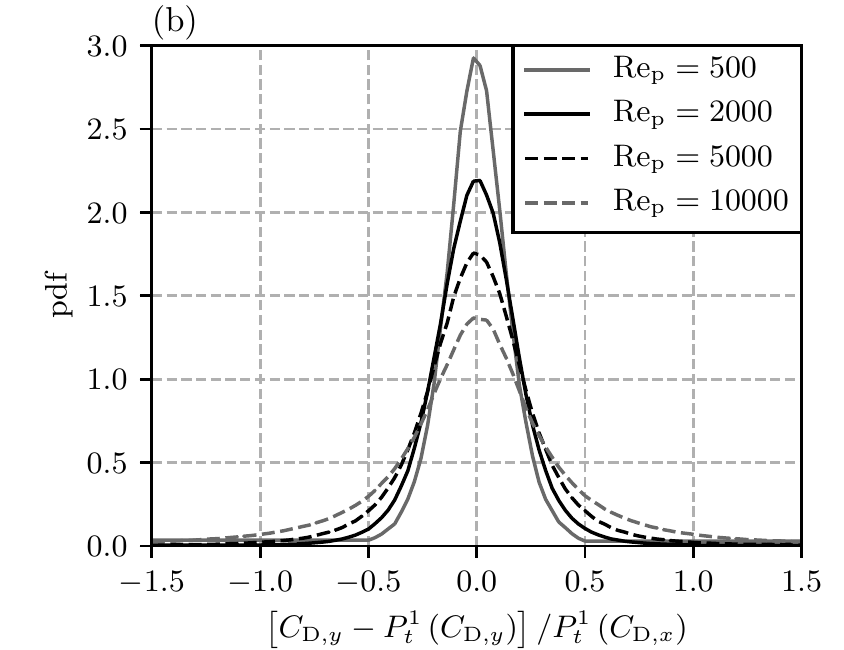}}
	\caption{Distribution of instantaneous drag coefficients for $t>1.875\tauL$. (a) Streamwise drag coefficient. (b) Spanwise drag coefficient. The linear trend with $t$ for each particle, denoted $P_t^1(C_{\mathrm{D}})$ has been subtracted.}
	\label{fig:cdx-instantaneous-distribution}
\end{figure*}

\begin{figure*}
	\centering
	\includegraphics[]{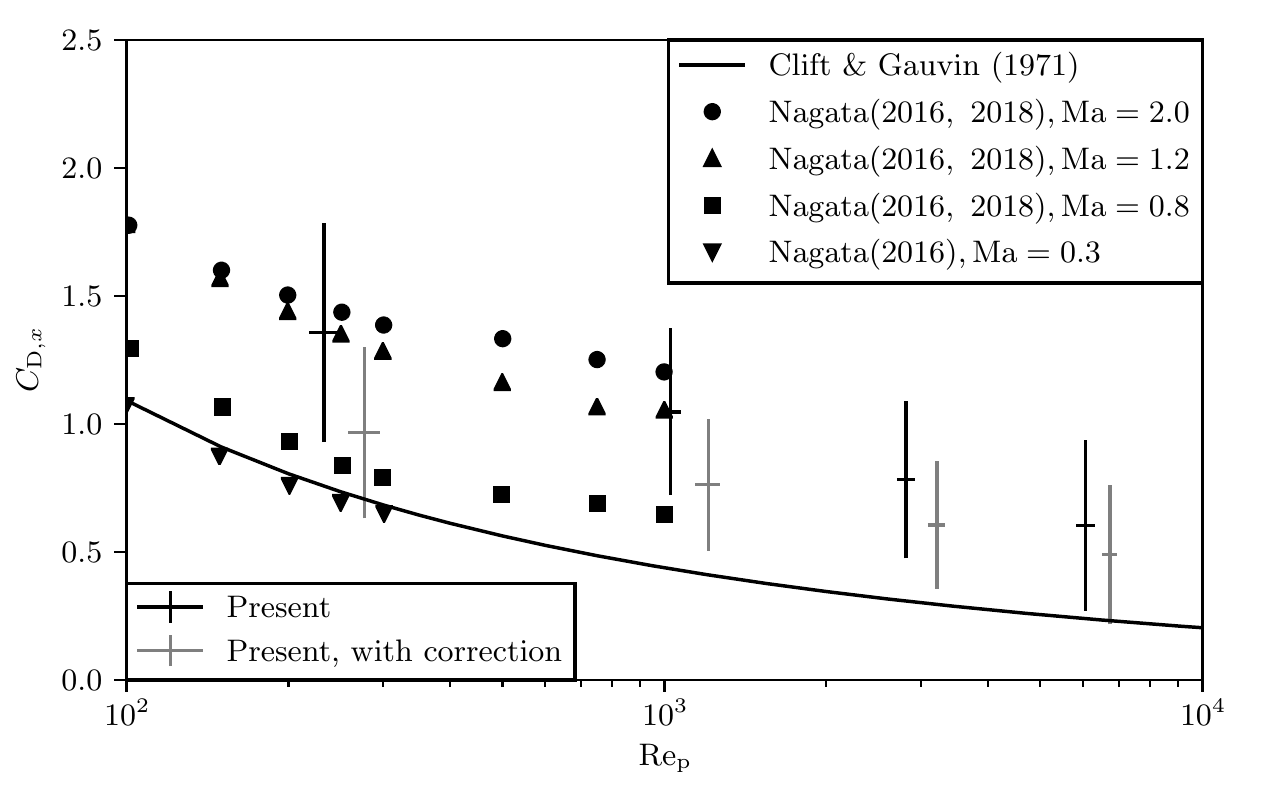}
	\caption{Average drag coefficient at $t=3.75\tauL$ for various $\Rep$, based on local (volume averaged) flow properties. The standard deviation in $\Rep$ and $\CDx$ are indicated by the horizontal and vertical error bars, respectively. The results of \citet{nagata2016,nagata2018} as well as the drag correlation of \citet{clift1971} are also shown. }
	\label{fig:late_cd_vs_rep}
\end{figure*}

Accurate description of particle forces is of key importance in dispersed flow simulations. In dense particle suspensions, the local particle configuration can lead to large variations in the directions and magnitudes of the particle forces. The maximal drag for each particle occurs during its interaction with the shock wave. The behaviour of maximal drag forces has been examined in inviscid simulations by \citet{mehta2018,mehta2019}. We compare the peak streamwise drag forces in the current simulations to their inviscid simulations in \cref{fig:peak_cdx_distribution}. The figure shows the trend in average peak streamwise drag forces with downstream distance and the distribution of peak streamwise drag forces. Note that for the results shown in \cref{fig:peak_cdx_distribution}, the drag forces are presented as a drag coefficient using $\rho_\mathrm{IS}$ and $u_\mathrm{IS}$. In \cref{fig:peak_cdx_distribution,fig:cdx-mean-distribution,fig:cdx-instantaneous-distribution,fig:late_cd_vs_rep}, subscripts $x$ and $y$ are used to indicate streamwise or spanwise drag coefficients, respectively. The shock wave Mach number is $2.6$ in the current simulations and $3$ in the inviscid simulations. We thus expect slightly higher drag coefficients in our simulations. The lower Reynolds number simulations show higher peak drag forces, and this is to be expected because the contribution from viscous forces increases. It can also be seen that the slope is steeper for the lower Reynolds numbers. This is due to the stronger shock wave attenuation occurring for lower Reynolds numbers. The distribution of the deviation of peak drag forces from the linear trend with $x$ is also shown in \cref{fig:peak_cdx_distribution}. The results are in excellent agreement with the inviscid simulations. It can also be seen that the distribution only has an inconsequential dependence on Reynolds number. This means that the peak drag variation is controlled by processes that are almost unaffected by viscosity, such as shock wave focusing and diffraction. 

After the initial shock-induced transient, the importance of viscosity increases. It is therefore of interest to investigate how the distributions of instantaneous and time-averaged drag coefficients depend on Reynolds number. \Cref{fig:cdx-mean-distribution} shows the distribution of the drag coefficients averaged over $1.875 \leq t/\tauL \leq 3.75$. Starting at $t=1.875\tauL$ ensures that any shock-related transients have decayed. The linear trend with $x$ has been subtracted from the drag coefficients and they are also normalized with that function. For the spanwise distributions, we have  normalized the drag coefficients based on the linear trend of the streamwise drag coefficients. The highest Reynolds number has a wider distribution of streamwise drag coefficients than the lower Reynolds numbers. The other three are very similar, but it seems that the distribution for $\Rep=2000$ is slightly higher and narrower than those for  $\Rep=500$ and 5000. The distributions are slightly skewed. This is most pronounced for $\Rep=10000$. The spanwise distributions are more similar for all $\Rep$. The highest Reynolds number has the widest distribution, and the distributions for the two middle $\Rep$ are almost equal.

\Cref{fig:cdx-instantaneous-distribution} shows the distribution of the deviation of the instantaneous drag coefficients from a linear trend in time. For each particle, the linear trend in time is also used for normalization. Since the trend in time has been subtracted, these distributions reflect the magnitude and frequency of temporal oscillations. The trend with $\Rep$ is similar for both the streamwise and spanwise drag-coefficients, where higher $\Rep$ result in wider distributions. Note that the spanwise drag coefficient  distribution is about twice as wide as that of the streamwise drag coefficient. The trends are more pronounced for the distribution of instantaneous than mean drag coefficients.

Both mean flow and instantaneous drag coefficient distributions have a dependency on particle Reynolds number. There are separate implications for drag modeling from each of these distributions. The distributions of time-averaged drag coefficients imply that drag models should incorporate terms that depend on the local particle configuration. These models can be based on the relative position of nearby particles or on the particle density. The PIEP model developed in \citet{akiki2017} is an example of the former model type. In the latter model type, the distributions shown in \cref{fig:cdx-mean-distribution} can be used directly by imposing random fluctuations drawn from these distributions. The results here also imply that the importance of including such distributions increases with Reynolds number since the distribution is widest at $\Rep=10000$. 

The distribution of instantaneous drag coefficients is a result of temporal flow fluctuations, and these have strong Reynolds number dependencies. It is likely that a part of these distributions is related to the shedding of vortices from both the particle itself and the nearby particles. The variation imposed by vortex shedding from nearby particles could be modeled based on the local particle configuration, although this is likely quite challenging. In addition to the trends with Reynolds number observed here, the distributions are likely dependent on the particle volume fraction. Investigating this effect is a topic for future studies.

\Cref{fig:late_cd_vs_rep} shows the streamwise drag coefficient at $t=3.75\tauL$. The figure also shows the drag coefficients where a modified flow velocity, defined as
\begin{equation}
u_\mathrm{free}=\favg{u}\frac{\alpha}{\alpha-\alps},
\end{equation}
is used to calculate the drag coefficient. The modified velocity was suggested in \citet{osnes2019} and attempts to correct for the contributions of the circulation regions behind the particles, which shift the average velocity away from the "free" velocity within the particle layer. Also shown in the figure are the single-particle drag coefficients obtained by \citet{nagata2016,nagata2018} and the drag correlation by \citet{clift1971}. For reference, the average Mach numbers within the particle cloud are in the range $0.3-0.7$, c.f. \cref{fig:flow_field_time_series_2}. Here, the lower Mach number is for $\Rep=500$ and the highest Mach number is for $\Rep=10000$, where the velocity correction has also been used for the Mach number. Even with the correction, which reduces the drag coefficients, the current simulations result in significantly higher drag coefficients than predicted by models based on single-particle results. The single-particle simulation results of \citet{nagata2016,nagata2018} only cover the lower two Reynolds numbers considered here. However, if the trend from these single particle simulation results can be extrapolated, it seems that the deviation is larger for low particle Reynolds numbers than high.

Other studies have found that confinement has a strong effect on the drag on single particles \citep{achenbach1974,akutsu1977,yeung2009,krishnan2010}. For $\Rep$ up to about 300, \citet{akutsu1977} found that a $30\%$ area blockage increased the drag coefficient by more than a factor of two. At $\Rep$ above $3\times10^4$, \citet{achenbach1974} found a $10\%$ increase for $50\%$ blockage. We expect that a similar effect occurs due to the presence of nearby particles. If a correction for this is combined with the corrections due to Mach number and separation volume, it is likely that the average drag coefficients will approach the values seen in isolated particle studies. 

\section{Concluding remarks\label{sec:concludingremarks}}

This study has investigated the propagation of a Mach 2.6 shock wave through particle clouds with particle volume fraction 0.1. Particle resolved large-eddy simulations were utilized, and the particle Reynolds number, based on the incident shock state, was varied between 500 and 10000.  

The results show that the shock wave attenuation increases with decreasing particle Reynolds number, but the effect is small over this Reynolds number range. The strength of the reflected shock wave increases with decreasing Reynolds number. This results in considerable changes in the flow quantities upstream of and inside the particle layer, i.e., higher pressures, lower velocities and lower Mach numbers with reduced particle Reynolds numbers. 

We examined the development of fluctuations from the volume averaged flow quantities within the particle layer. The velocity fluctuation intensity has a non-monotone dependency on particle Reynolds number. The most intense velocity fluctuations were found at $\Rep=2000$. Streamwise velocity fluctuations are located in the particle wakes, and most importantly in the circulation region behind each particle. Spanwise fluctuations are located in the inter-particle regions, and consist of coherent regions that extend further than the size of single particles. As the Reynolds number increases, the sizes of these regions decrease. We find an approximate proportionality between the mean flow velocity and the flow fluctuations at late times. This relationship holds for both the streamwise and the spanwise fluctuations. Interestingly, the proportionality factor varies monotonously with $\Rep$ for the streamwise fluctuations, but not for the spanwise fluctuations. 

The distributions of particle forces were examined. Maximal drag forces and their spatial variation agree well with the inviscid simulations of \citet{mehta2018}. In the late-time flow field, we find that both the variation of time-averaged particle forces and the temporal variation increases with particle Reynolds number. The trend with particle Reynolds number is more pronounced for the temporal variation than the time-averaged distribution. These distributions provide a basis for development of particle force variation models, that can be used in simplified dispersed flow models. 

The particle drag coefficients within the particle cloud were found to be higher than isolated particle drag coefficients on average, and the largest deviation occurred for the lowest Reynolds number. We demonstrated that utilizing a correction to the mean flow velocity yields drag coefficients closer to the single-particle results. A combination of this correction with drag correlations that incorporate Mach number and flow blockage effects, can possibly reproduce the observed average drag coefficients in shock-wave particle cloud interactions.

\end{document}